\documentclass[12pt,manuscript]{aastex}

\slugcomment{Submitted to AAS Publishing}

\def\kms{\ifmmode{\rm km\thinspace s^{-1}}\else km\thinspace s$^{-1}$\fi}
\def\ms{\ifmmode{\rm m\thinspace s^{-1}}\else m\thinspace s$^{-1}$\fi}

\shortauthors{Howell, et al.}
\shorttitle{Rapidly Rotating Stars}

\begin{document}

\title{Rapidly Rotating, X-ray Bright Stars in the $Kepler$ Field}

\author{
Steve~B.~Howell\altaffilmark{1} \\
NASA Ames Research Center, Moffett Field, CA 94035, USA \\
Elena~Mason \\
INAF-OATS, Via G. B. Tiepolo 11, 34143, Trieste, IT \\
Padi Boyd\altaffilmark{1} \\
NASA Goddard Space Flight Center, Greenbelt, MD 20771, USA \\
Krista Lynne Smith\altaffilmark{1} \\
Department of Astronomy, University of Maryland College Park, USA and \\
NASA Goddard Space Flight Center, Greenbelt, MD 20771, USA\\
Dawn M. Gelino\altaffilmark{1} \\
NASA Exoplanet Science Institute, Caltech, Pasadena, CA 91125, USA \\
}

\altaffiltext{1}{Visiting Astronomer, Mt. Palomar Observatory 200" Hale Telescope}

\begin{abstract} 
We present $Kepler$ light curves and optical spectroscopy of twenty X-ray bright stars located in the $Kepler$ field of view. 
The stars, spectral type F-K, show evidence for rapid rotation including chromospheric activity 100 times or more above the Sun at maximum and flaring behavior in their light curves. Eighteen of our objects appear to be (sub)giants and may belong to the class of FK Com variables, 
that is evolved rapidly spinning single stars with no excretion disk and high levels of chromospheric activity. 
Such stars are rare and 
are likely the result of W UMa binary mergers, a process believed to produce the FK Com class of variable and their descendants. The FK Com stage, including the presence of an excretion disk, is short-lived but leads to longer-lived stages consisting of single, rapidly rotating evolved  (sub)giants with high levels of stellar activity.
\end{abstract}


\section{Introduction} 

The NASA $Kepler$ mission (Borucki et al., 2010) 
was launched in 2009 and completed four years of photometric observation
of over 150,000, mainly late-type stars in a single field of view. The $\sim$100 sq. degree field was located between the constellations of 
Cygnus and Lyra, including a part of Draco as well.  A number of photometric surveys of the $Kepler$ field of view were conducted in order to exploit the information available from the mission complementing, for example, the $Kepler$ light curves with ground-based multi-band photometric characterization of the sources or spectroscopic observations (e.g.  Everett et al. 2012, Greiss et al. 2012, Huber et al., 2014)
 
Our team conducted the $Kepler$-Swift Active Galaxies and Stars survey (KSwAGS) using the Swift X-ray
Telescope (XRT), which operates in the 0.2-10 keV range and includes co-aligned UV-optical telescope (UVOT). This survey of the $Kepler$ FOV covered about 6 square degrees imaging a strip of sky roughly perpendicular to the galactic plane in order to sample a range
of galactic latitude. The survey produced X-ray and simultaneous UV
information (for most sources); about 30\% of
the X-ray sources do not have UVOT coverage because of the smaller UVOT FOV. 
Details of the KSwAGS survey can be found in Smith et al. (2015).

Within the KSwAGS survey we found over 90 sources with significant f$_{x}$/f$_{v}$ values; $>$100 times the Sun at maximum, i.e. $log$(f$_{x}$/f$_{v}$)=--6.  Of these, 60 had UV counterparts that were matched to sources in the $Kepler$ Input Catalog sources (KIC; Brown et al. 2011). 
Initial spectroscopic observations were obtained for 30 of the brightest X-ray sources yielding identifications including a 
handful of AGN discussed in Smith et al. (2015), two active M stars (KSw 64 and KSw 84, not discussed further here) and the 20 stars discussed in this paper.   

In this treatise, we make a case that the majority of these f$_{x}$/f$_{v}$ bright stars are rapidly rotating, single, evolved stars, i.e., candidate FK Com or FK Com descendants. We present medium resolution optical spectra, which we used for spectral and luminosity classification, and analysis of $Kepler$ light curves which we used to determine periods and, for those stars whose light modulations are consistent with variations induced by rotation, stellar radii.

\section{Observations}


The optical counterpart for each of our KSwAGS X-ray sources was identified in
the Kepler Input Catalogue (KIC) based on a coordinate match with the UVOT image
measured positions, when available, or the detected X-ray source position coordinates
(Smith et al., 2015). UVOT images can localize a source to within 1.4 arc seconds when astrometrically
corrected (Goad et al., 2007) and reliably provide positions to within 0.5 arc seconds (Breeveld et al., 2010). 
In all but
one case, the UVOT coordinates provided a match to a KIC source coordinate to within 2Ó. The one exception,
based on only the X-ray coordinate (KSw 16), matched a bright (V=12.6) KIC source to
within 4Ó. The fact that Ksw 16 displays similar spectral and timing characteristics as the other
stars in our sample lends confidence that it is the correct identification.
Once a KIC star identification was obtained, we could associate the source with
its Kepler light curve (when available) as well as optical magnitude information and any
published literature references. We provide the Kepler identification number (KIC number)
and KSwAGS source (KSw number) in Table 1 and refer to the sources throughout the
paper by their KSw number.

\subsection{Photometric Observations with $Kepler$}

$Kepler$ photometric light curve data was obtained during the nominal mission (2009-2013) and downloaded from the spacecraft
in three month sets called quarters. During each quarter, the spacecraft remained in the same orientation with respect to the stellar images, that is the stars fell onto the same pixel locations in the focal plane. After each quarter, the
spacecraft would rotate 90 degrees in order to keep the solar panels facing the Sun, thus placing the
entire field onto different CCD locations within the rotationally symmetric focal plane. The $Kepler$ quarters are named
Q1,Q2, Q3,...Q17, with Quarter 0 being an initial short 10-day segment.
 
$Kepler$ light curves provide photometric time series observations with two possible integration times; 30 minutes (long
cadence) and 1 minute (short cadence). All but three of our KSw stars had long cadence monitoring. A much smaller subset (the five brightest stars) were observed in short cadence mode as well for one quarter each. Table 1 provides an observing log for the $Kepler$ data and lists our program stars, giving the $Kepler$-Swift survey number (Smith et al. 2015) as well as the KIC identification number. Table 1 also lists the quarters in which
each star was observed in either long or short cadence. Stars KSw 54, 66, 76 are located in the $Kepler$ field of view but did not have light curves recorded by $Kepler$. 

We downloaded the simple aperture photometry (SAP) $Kepler$ light curves used in this paper from the MAST archive\footnote{https://archive.stsci.edu/kepler/} and did no additional processing of them prior to our period searching. Figures 1-3 show one representative quarter of $Kepler$ long cadence light curve observations for each of our stars. The photometric precision for each  of the plotted light curves, as derived via the $Kepler$ pipeline\footnote{See the Kepler instrument Handbook: http://archive.stsci.edu/kepler/manuals/KSCI-19033-001.pdf}, is listed in Table 1. The quarters were arbitrarily chosen to allow the stars to be grouped in the figures. The KSw identification number is given in the figures.  The light curves for all stars are variable and most present quite complex behaviors due to spot modulations and differential rotation, not the simple robust single orbital period a binary star would reveal. We will discuss period searches and whether the light is due to rotation, binarity or pulsations in \S3.2. 

\begin{figure}
\label{f1}
\includegraphics[angle=0,scale=0.75,keepaspectratio=true]{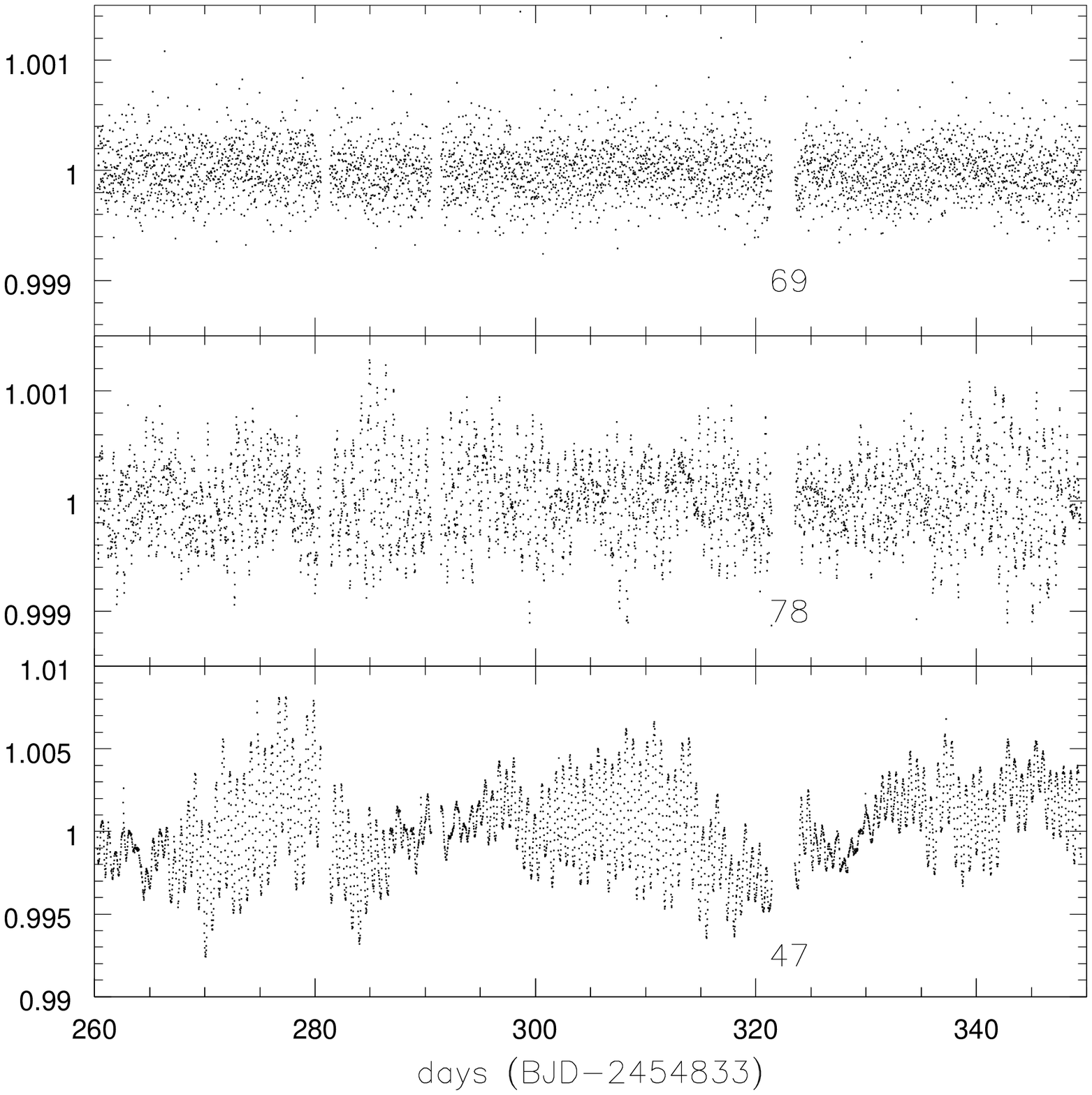}
\caption{Quarter 3 long cadence $Kepler$ light curves for KSw stars 47, 69, and 78. The KSw identification number for each star is indicated nearby the light curve. The y-axis is the median-normalized counts. 
}
\end{figure}

\begin{figure}
\label{f2}
\includegraphics[angle=0,scale=0.75,keepaspectratio=true]{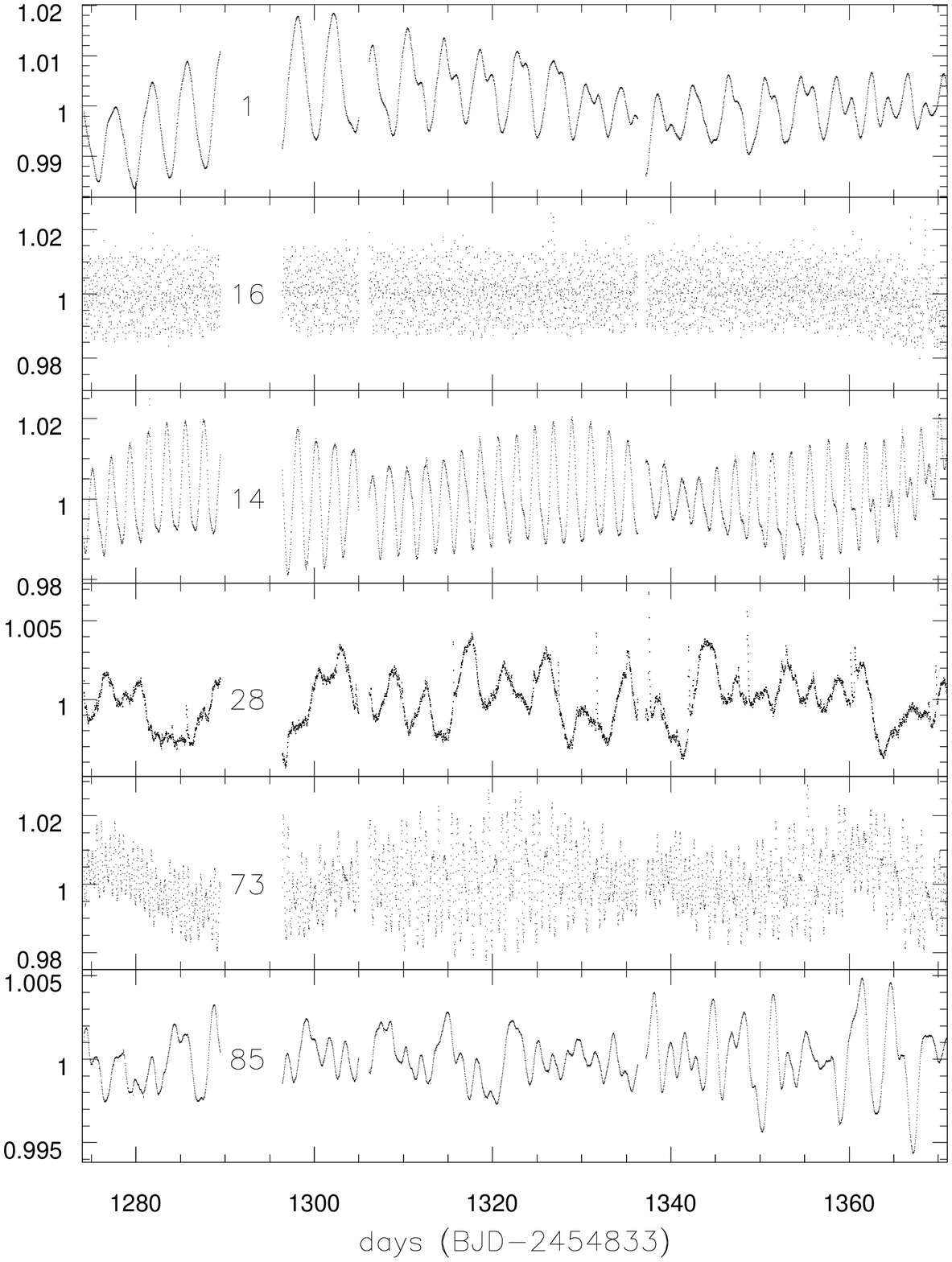}
\caption{Quarter 14 long cadence $Kepler$ light curves for KSw stars 1, 14, 16, 28, 73, and 85. The KSw identification number for each star is indicated nearby the light curve. The y-axis is the median-normalized counts. 
}
\end{figure}

\begin{figure}
\includegraphics[angle=0,scale=0.75,keepaspectratio=true]{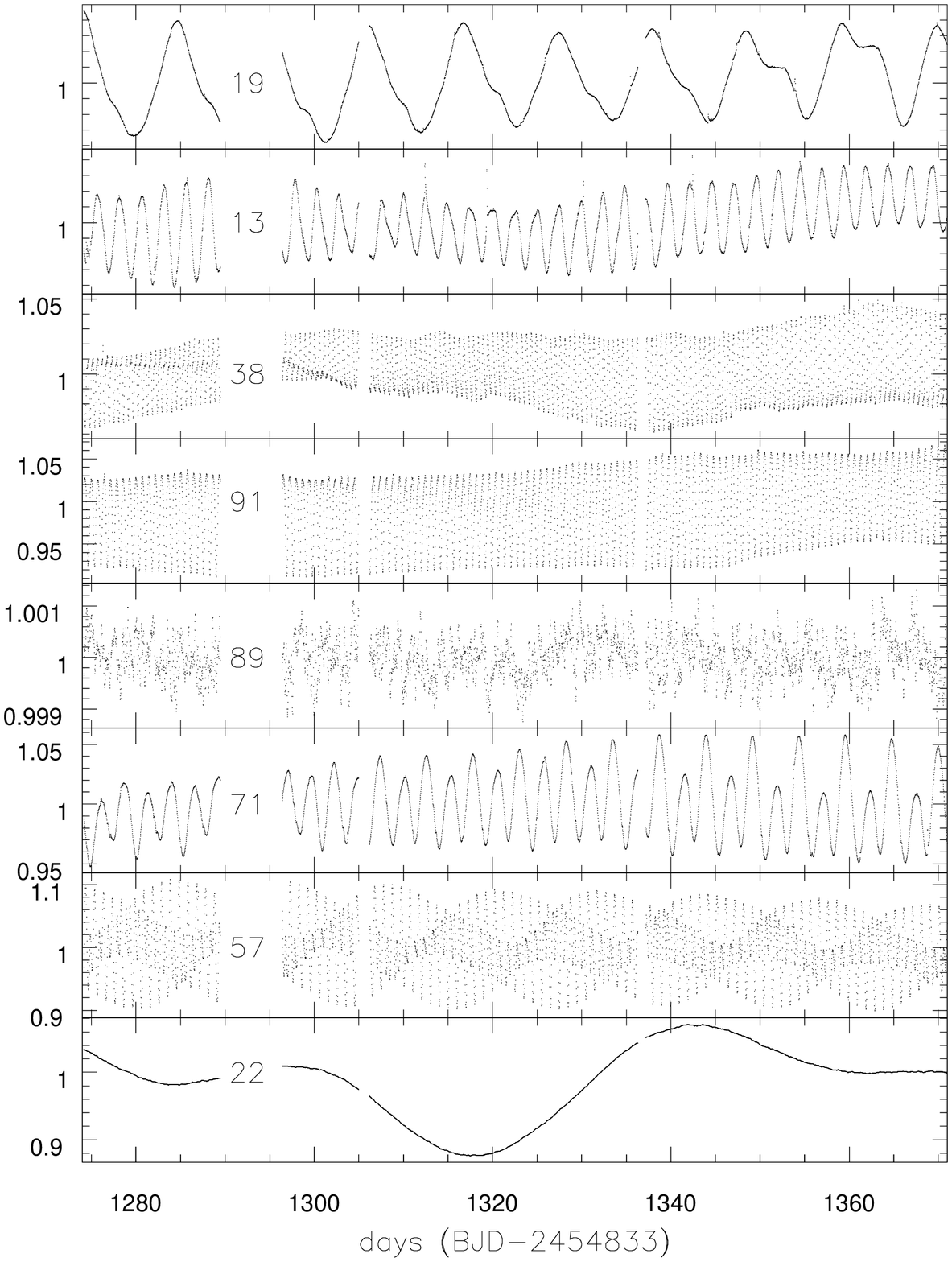}
\caption{Quarter 14 long cadence $Kepler$ light curves for KSw stars 13, 19, 22, 38, 57, 71, 89 and 91. The KSw identification number for each star is indicated nearby the light curve. The y-axis is the median-normalized counts. 
}
\end{figure}

\subsection{Optical Spectroscopy with Hale 200" Telescope}

During August 2014 we obtained spectral observations of the 20 sources which are
discussed in this paper. The observations were made with the double beam spectrograph (DBSP)
attached to the Mt. Palomar 200-inch Hale telescope. 
Dichroic filter D-55 was used to split light between the
blue and red arms of the spectrograph. 
The blue arm used a 1200 l/mm grating providing R$\sim$7700 and covered
1500\AA~of spectrum. The red arm used a 1200 l/mm grating providing R$\sim$10,000 and covered
only 670\AA. The slit width was set to 1 arcsec and the usual procedure of observing
spectrophotometric stars and arc lamps was adhered to. Red spectra were wavelength calibrated
with a HeNeAr lamp while the blue arm used a FeAr lamp. The nights were clear and provided stable
seeing near 0.9-1 arcsec. Table 2 presents our spectroscopic observing log and lists the $V$ magnitude for each source. Integration times were chosen to provide a S/N of 35-60 per resolution element near the central wavelength of each beam.

The observing procedure was as follows. The telescope was pointed and set to autoguide and each
pointing began with an exposure of an arc lamp for wavelength calibrations. Following
that, the star was observed. Spectrophotometric calibration stars were observed occasionally throughout
the night to allow the
fluxes to be placed on a relative scale. Additional calibration data consisting of bias frames and 
quartz lamp flat field exposures were obtained at the start of each night.

Data reduction used the {\it twodspec} IRAF packages for performing initial image calibration and
spectral extraction and the
{\it onedspec} package for final calibration of the spectra. A sensitivity function is
found for each night based on the ratio of the standard star to its standard curve in the IRAF
database of Kitt Peak IRS standards and used to correct the science targets and supply a relative
flux level. The comparison lamp spectra are used to determine wavelength as a function of
columns in order to resample the spectra to a linear wavelength scale set to closely match the
sampling of the two-dimensional spectra. 

Figures 4-7 show the blue and red spectra respectively for our stars including expanded region views of the Ca II H\&K and 
H$\alpha$ lines.  The spectra were all boxcar smoothed by 3 for presentation. The KSw number is listed on each spectral plot and the blue and red spectra are displayed in the same order for each set. Figures 4 and 5 show the early type stars in our sample (A through early G) and we note that none of their spectra show H$\alpha$ in emission while the spectra of the later stars in Fig. 5 begin to show weak Ca II emission core reversals. Figures 6 and 7 show the later type stars in our sample (mid-G to late K) and here we note that all these spectra show Ca II emission core reversals and often complex H$\alpha$ profiles. We see no evidence in any of the spectra for a binary companion.

\begin{figure*}
\includegraphics[angle=0,scale=0.75,keepaspectratio=true]{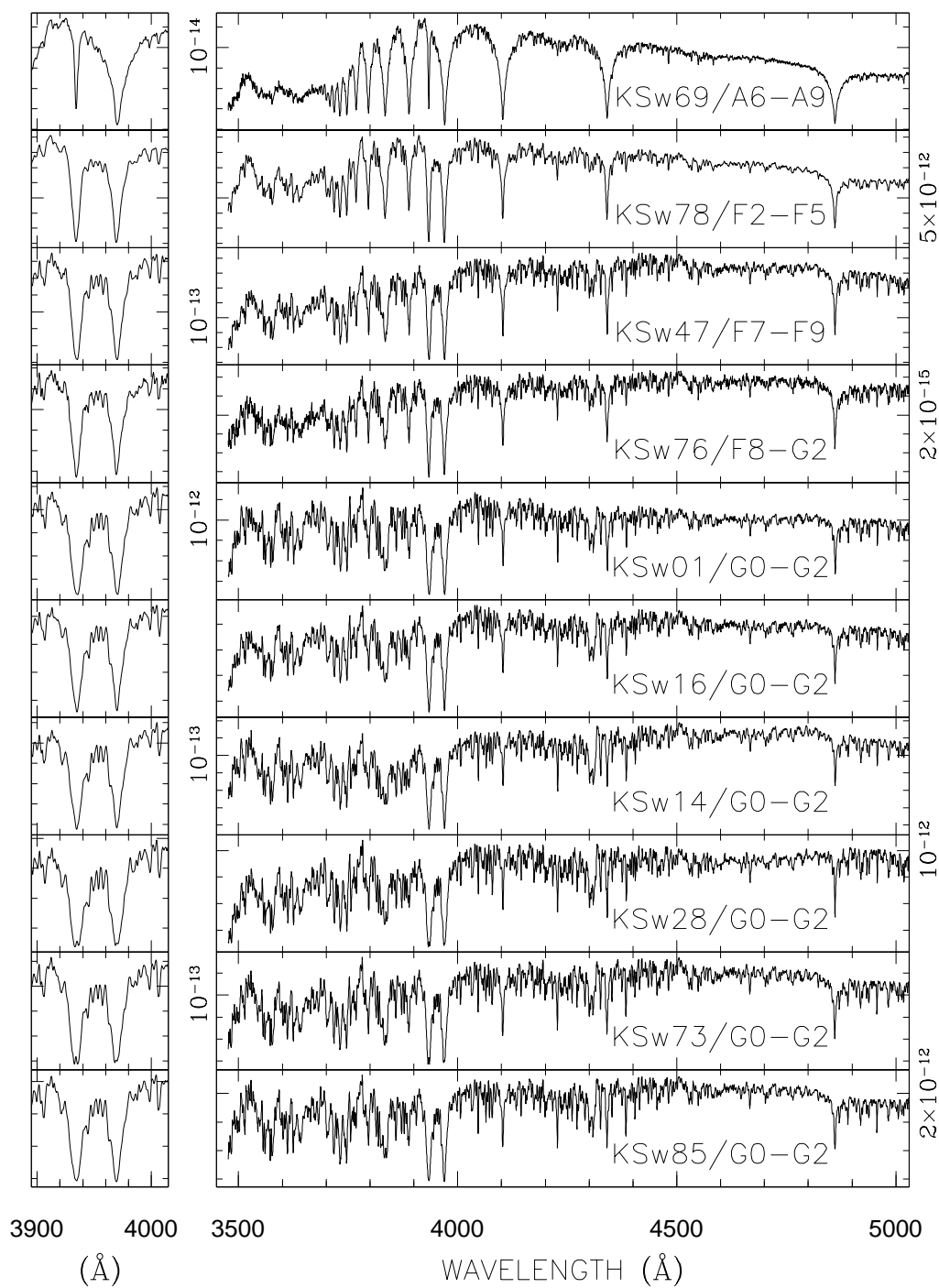}
\caption{Blue spectra of half our star sample presented in the same order as in Figure 5. The Y-axis is relative flux in ergs/cm$^2$/sec/\AA~and the left zoomed panels illustrate an expanded view of the Ca II H\&K region.
}
\end{figure*}

\begin{figure*}
\includegraphics[angle=0,scale=0.75,keepaspectratio=true]{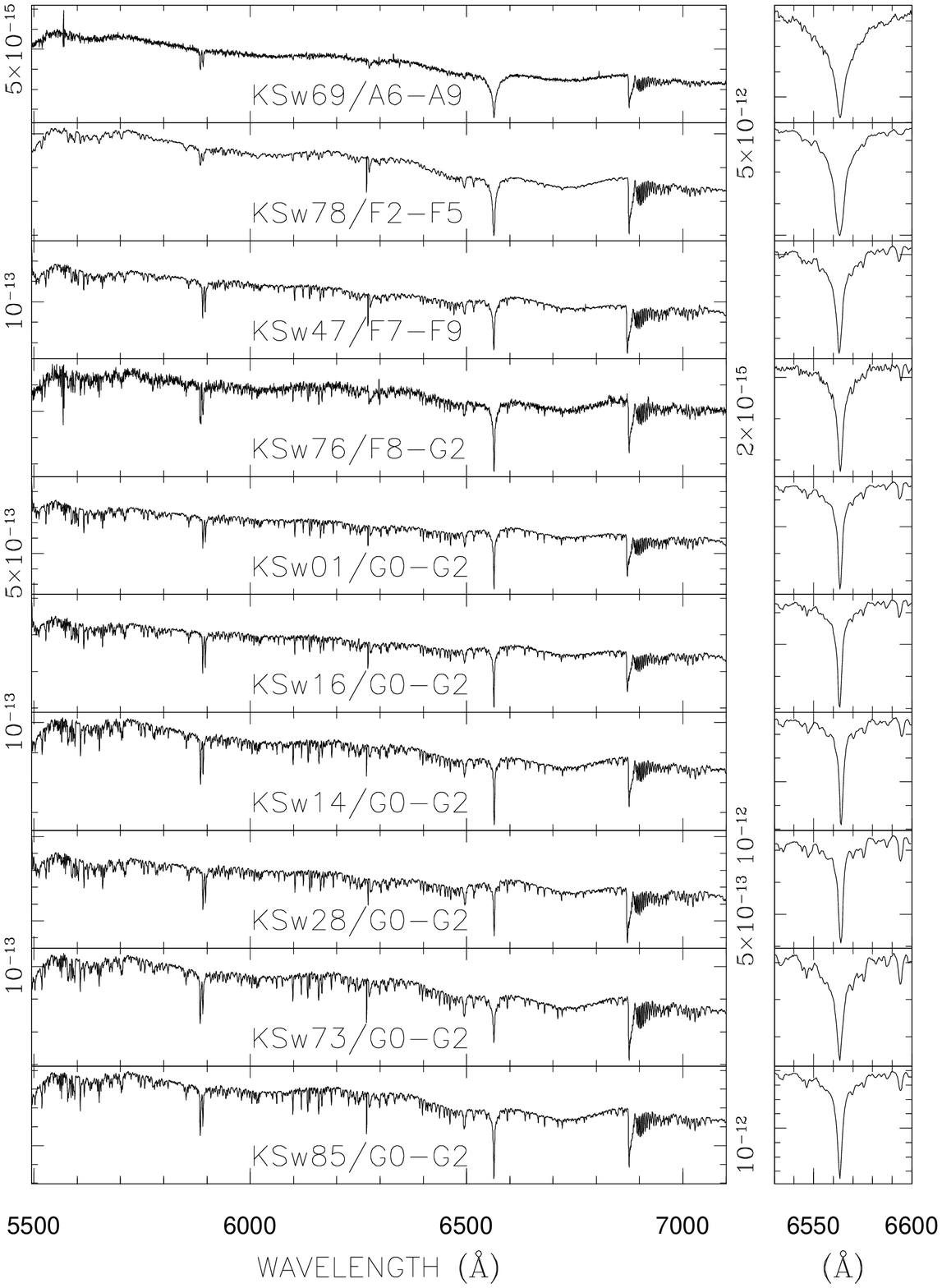}
\caption{Red spectra of half our star sample corresponding to the Blue spectra in Figure 4. The Y-axis is relative flux in ergs/cm$^2$/sec/\AA~and the right zoomed panels illustrate an expanded view of the H$\alpha$ region.
}
\end{figure*}
\begin{figure*}
\includegraphics[angle=0,scale=0.75,keepaspectratio=true]{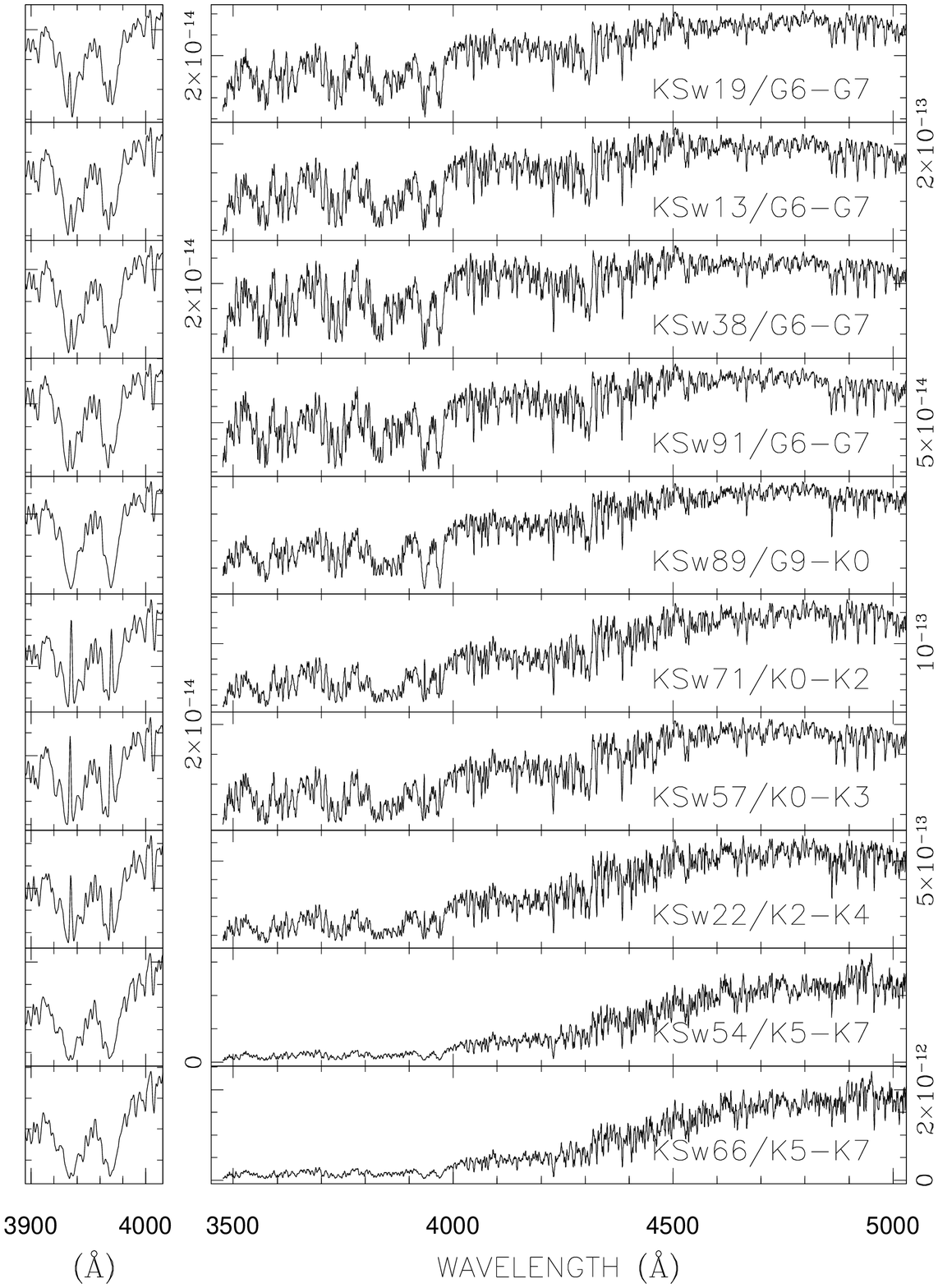}
\caption{Blue spectra of half our star sample presented in the same order as in Figure 7. The Y-axis is relative flux in ergs/cm$^2$/sec/\AA~and the left zoomed panels illustrate an expanded view of the Ca II H\&K region.
}
\end{figure*}

\begin{figure*}
\includegraphics[angle=0,scale=0.75,keepaspectratio=true]{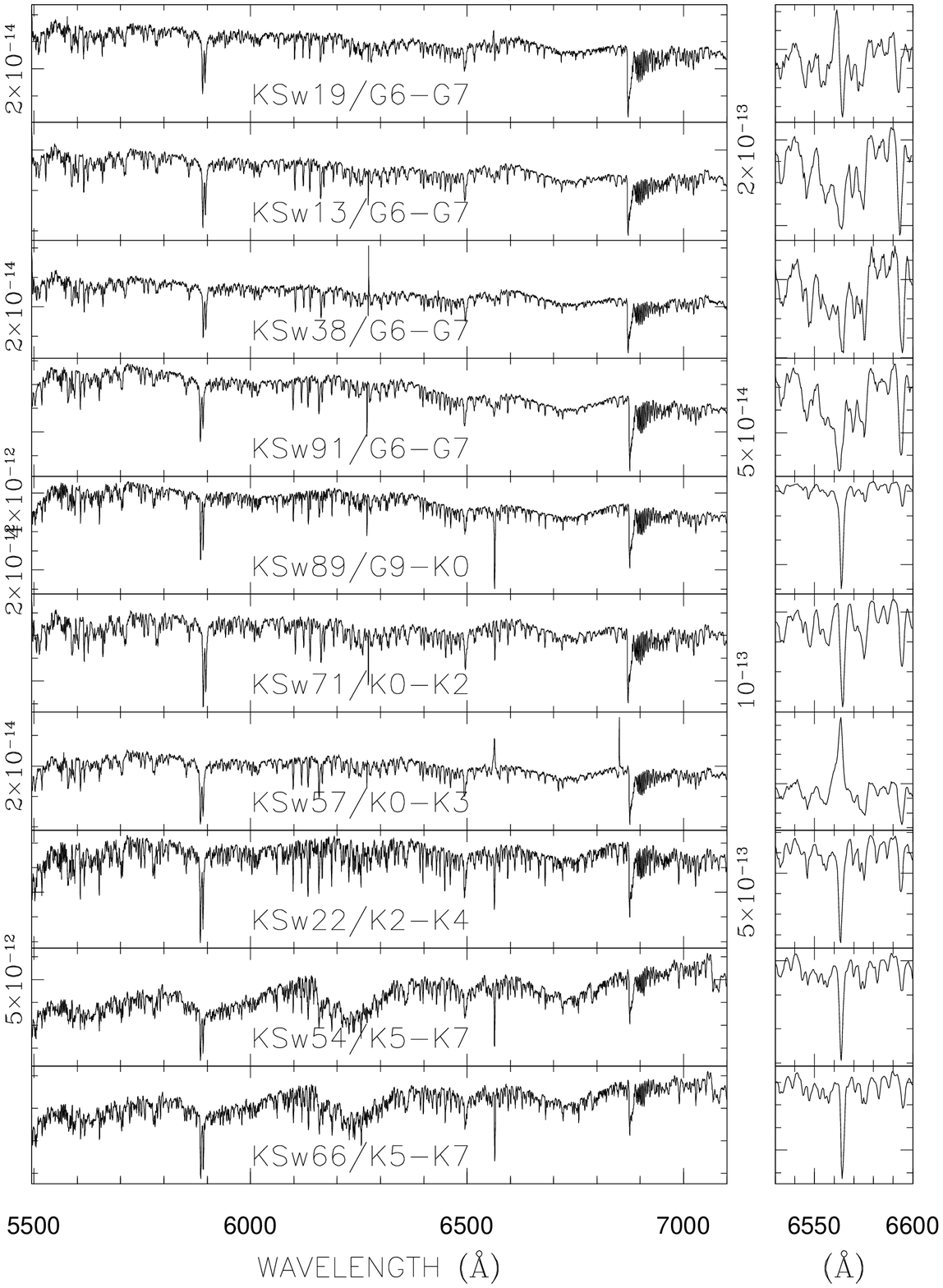}
\caption{Red spectra of half our star sample corresponding to the Blue spectra in Figure 6. The Y-axis is relative flux in ergs/cm$^2$/sec/\AA~and the right zoomed panels illustrate an expanded view of the H$\alpha$ region.
}
\end{figure*}

\section{Analysis}
\subsection{Spectral Type, Luminosity Class,  and $v~sin~i$ Measurement}

Table 3 presents our determined spectral type and luminosity class for each star. We provide a description of the usual active chromosphere indicators, our $v sin i$ measurements, the x-ray to optical ratio (from Smith et al. 2015), and a yes/no flag to indicate 
if the stars show flares in their $Kepler$ light curves. We include FK Com in this table for reference.

Each star's spectral type was determined by relative comparison to MK standard stars as presented and discussed in the atlas of Jacoby et al. (1984) and the study of Gray and Corbally (2009). For the luminosity class determination, we primarily used the blue spectral region, in particular the strength of the CN and g-Band absorptions, the Balmer-jump, and the general continuum shape.  In addition to those gross features, we inspected the relative line strengths of the Sr\,{\sc ii} 4077 line over Fe\,{\sc i} 4046 and of Y\,{\sc ii} 4376 over Fe\,{\sc i} 4383 as well as the line wings of Ca I and Fe I lines in comparison with LTE solar metallically model spectra broadened to our spectral resolution. Our final classification in Table~3 results from a decisional mean of the various methods. 
Although we compared our spectra to high resolution spectral models, we did not attempt spectral fitting with such models as this was found to be prone to degeneracy due to the moderate spectral resolution of our data.  
The stars appear to be F-K spectral type and slightly to more evolved, mainly residing in luminosity classes of IV or III. We will see below that our spectral classifications  are in good agreement with previous literature studies of some of our stars.

We used line-width measurements as a proxies for $v sin i$ following the prescriptions of Shajn and Struve (1929), as modified and improved by Aller (1963), Slettebak et al. (1975), and Gray (1989).  IRAF's onedspec line fitting routines were used to measure the absorption line profiles used to yield the $v sin i$  values. For our line measurements we used lines in both the red and blue arms such as Ca\,{\sc i} and Fe\,{\sc i} as well as other low ionization excited transitions which appear quite isolated in the high resolution spectrum of the Sun (most of the stars in our sample are near solar type, i.e. G2V). 
Given that our spectral resolution is only $\sim$60 ($\sim$70) km/s in the red (blue) arms, all $v sin i$ measurements of the order of 60-70 km/s shall be considered unresolved, indicating a defined lower limit to our measurement ability. In addition, due to the fact that a rotationally broadened spectral line is not well approximated by a Gaussian or even a Gaussian + Lorentzian line profile fit, we expect large uncertainties in the $v sin i$ values (due to line blending for example) of order 20\%. Admittedly, the 
moderate resolution spectra we have available are not suitable to reach typical stellar $v sin i$ values, however, our moderate spectral resolution does allow us to measure the large 
$v sin i$ values present in a number of our stars.

The X-ray to optical ratio, f$_{x}$/f$_{v}$, listed in Table 3 are taken directly from Smith et al., (2015). The values range from -4.7 to -1.2 dex. For comparison, the Sun at solar maximum has an X-ray flux of 5e27 ergs/sec and a log f$_{x}$/f$_{v}$ value of -6 while FK Com has 
log f$_{x}$/f$_{v}$ = -4.0 to -3.5. Therefore, our smallest value for log f$_{x}$/f$_{v}$ is about 100 times greater than that of the Sun at solar maximum, but well in line with the levels seen in the FK Com stars - stars with very active chromospheres. 
The determined $v sin i$ values tend to group in two ranges independent of spectral type.  About half of the stars are $\le$65-85 km/sec (that is, near or fully unresolved) with the rest being 100 km/sec or more, the latter group tending to be the stars with the highest f$_{x}$/f$_{v}$ ratios. The smallest x-ray/optical flux ratios occur for the two coolest stars in our sample,  
KSw 54 and KSw 66.
About half of the stars show a direct spectral indication of the presence of a chromosphere as evidenced by the Ca II H\& K line (emission cores) and many of the spectra show H$\alpha$ emission as well. Taken together, these special signatures are typically indicators that high levels of stellar activity will be present in these late-type stars. Rapid rotation and its relationship to chromospherically active single giants, including FK Com stars, has been investigated by  Fekel and Balachandran (1993).

\subsection{Light Curve Period Search}

The $Kepler$ light curves for the stars under study here generally show complex behaviors and only a few reveal a single well defined period.
For many, the amplitude of the modulation changes throughout, likely due to star spots and their movements, lowering the robustness and ease of period fitting. The total 
peak-to-peak amplitude in the light curves are small, only a few percent (see Figs. 1-3). We performed period search analysis for each light curve using the Lomb-Scargle  technique as implemented at the NExScI Exoplanet Archive as a web-based application\footnote{http://exoplanetarchive.ipac.caltech.edu}. 
The results of our period search are listed in Table 4 where we assign a likely type of period (e.g. rotational or pulsational) to each star based on the power spectrum properties, the value of the period itself, and period type assignments following methodology 
described in De Medeiros et al. (2013).
Very complex light curves or those showing only weak periods are denoted by a ``:" in the table, revealing the uncertain nature of the period. 

The periods found and listed in Table 4 agree with the more detailed period searches for seven common stars (KSw 1,14,19, 28, 38, 85, 91) presented in the literature (see \S4). The periodograms showed generally broad peaks, Gaussian-like in shape but often with an asymmetric distribution. That type of peak in a power spectrum argues for a rotation period that is modulated by star spots, their changing aspect contributing to the peak breadth and changing shape over time, not one that suggests that the modulations are that of a very stable binary period.
We find evidence for a binary period in only one star, KSw 85, a conclusion supported by its sharp narrow periodogram peak and literature spectroscopy (see \S4.7).

For those stars where we were fairly confident about the rotational nature of their period, we made use of the determined 
$v sin i$ value above to calculate a lower limit to the star's radius. These values support our luminosity class determinations for the stars - that is, the stars are evolved beyond the main sequence and have radii placing them in the subgiant or giant regime.
  
\section{Our KSw stars in the literature}
Half of the stars in our sample have been studied in some manner by other $Kepler$ mission programs, especially as most  are relatively bright (Table 2).
For example, 
a number of our sample has already had their $Kepler$ light curve (or part of it) analyzed in detail (KSw 1, 14, 19, 38, 85, 91: see below) yielding rotational period determinations (Nielsen et al., 2013; McQuillan et al., 2014; Reinhold et al. 2013). Only three of our KSw stars have published high resolution spectroscopy with model fits (Molenda-Zackowicz et al 2013, Guillout et al. 2009) yielding accurate log g values. 
We summarize below the relevant literature focusing, in particular, on results related to rotational periods and stellar parameter determinations. We note that our determinations herein agree with these literature references for the stars we have in common.

\subsection{KSw 1}

The $Kepler$ light curve was searched for a rotational period both by Nielsen et al. (2013) and Mc Quillan et al. (2014). Nielsen et al. analyzed $Kepler$ Quarters 2 to 9, searching for periods in the range 1 to 100 days. They identified the most stable period as the period of rotation for the star, P$_{rot}$=4.098 d. Similarly, analyzing $Kepler$ Quarters 3 to 14 with an automated autocorrelation function, McQuillan et al. (2014) found P$_{rot}$=4.095$\pm$0.003 d. Our value of 4.13 d in Table~4 is a close match to both of these more detailed examinations.  

KSw 1 was also observed with high resolution spectroscopy by Molenda-Zackowicz et al. (2013) on three different occasions with HERMES and FRESCO spectrographs. From the modeling of the spectra, they determined the stellar parameters, T$_{eff}$, $log~g$, [Fe/H] and $v sin i$ and suggest that the star is a $\geq$6000 K dwarf with v\,sin$i\sim$13-15 km/s. Guillout et al. (2009) analyzing high resolution spectra centered on the H$\alpha$ and Li\,{\sc i} region find a similar temperature, a somewhat smaller log g, and somewhat larger $v sin i$ (20.1 km/s). These $v sin i$ values are consistent with our inability to resolve its $v sin i$ due to our much lower spectral resolution, while our luminosity class determination {V-IV}, is in good agreement with the high resolution spectral modeling. KSw 1 is not likely an FK Com candidate and the reason for its large X-ray to optical ratio is unknown at present. 

\subsection {KSw 13 and KSw 71} 
Pigulski et al (2009) discuss the strictly sinusoidal light curves of KSw 13 and KSw 71 for which they find periods of
P=2.4228 d and P=5.212 d, both in good agreement with our periods present herein.
While Pigulski et al.  suggest that the light curves are sinusoidal and thus due to pulsations, We note that for these two stars, their location in the H-R Diagram lies at the far red edge of  the region occupied 
by the $\delta$ Sct pulsating variables. These variables typically have short periods in the range of 0.02-0.3 day, thus, the long periods measured ($>$1 day) seem more consistent with rotation periods.

\subsection {KSw 14} 
The chromospherically active nature of this star was already reported by Balona (2015) who identified 5 flares in the $Kepler$ Quarter 3 light curve alone and determined a rotation period of 2.064 d. We find flares as well and our period (2.06 d) is in very good agreement with that of  Balona  (2015).

\subsection {KSw 19}
The KSw19 $Kepler$ light curve has been analyzed by McQuillan et al. (2014) and Reinhold et al. (2013), before us. They find rotational periods of 10.775$\pm$0.004 d and 10.9208$\pm$0.0151 d, respectively. Also for this star, Reinhold et al. find a close-by secondary period (9.9446$\pm$0.0492) which is interpreted as evidence of differential starspot rotation. Our primary period (10.74 d) is a close match with the determined rotation period but we do not detect a separate 9.9 day period. Our non-detection of this close period may be due to the broad periodogram peak of the 10 day period and our use the complete $Kepler$ data set not just one quarter. We did note a weaker second period of 5.34 d, about half the 10d period, possibly due to star spots located 180$^o$ from each other on the surface of the star. We note that Pigulski et al. (2009) too, report a period of 10.817 d for this star. They classify KSw 19 as a quasi-periodic variable. 

\subsection {KSw 28}
This star has had different portions of the $Kepler$ light curve analyzed by several authors searching for the rotational period. Nielsen et al. (2013) searched for periods in the range 1 to 100 days using Quarters 2 to 9. With the Lomb-Scargle periodogram they identified the most stable period they find as the rotation period, P$_{rot}$=9.836 d. McQuillan et al. (2014) analyzing $Kepler$ Quarters 3 to 14 with an automated autocorrelation function found P$_{rot}$=9.718$\pm$0.490 d. Reinhold et al. (2013) analyzing Quarter 3 $Kepler$ data for over 1000 active stars finds  P$_{rot}$=10.2308$\pm$0.0192 d as the primary period and a secondary period P=9.3198$\pm$0.0255 d, which they interpret as a signature of differential starspot rotation. Finally, Balona (2015) inspecting Quarters 0 to 12  for over 20000 $Kepler$ light curves searched for flares and identified KSw 28 as showing multiple flares. We determine KSw 28's rotation period to be P$_{rot}$=10.96 d and noted flares in the $Kepler$ light curves as well, 
both values in good agreement with previous literature studies. 

\subsection {KSw 47} 
McQuillan et al. (2014) find a rotation period of 2.667$\pm$0.578 d for this star in close agreement with our period listed in Table 4. 

\subsection {KSw 85}
This star was observed with high resolution spectroscopy by both Molenda-Zackowicz et al. (2013) and Guillout et al., (2009). These authors find somewhat different results, the former suggest an early G dwarf ($log~g$=4.37, T$_{eff}\sim$5900 K); while the latter identify it a double lined spectroscopic binary modeling the two components: $log~g$=4.37 and T$_{eff}\sim$4900 K for component A, and $log~g$=3.41 and T$_{eff}\sim$5600 K for component B. The two measures of   
v\,sin$i$ by these two studies yield 24.5 and 11.1 km/s, respectively, consistent with our unresolved value. KSw 85's rotational period was determined by Nielsen et al. (2013; P=3.938 d), Reinhold et al. (2013; P= 4.4614$\pm$0.0064 d with a secondary period of 3.4298$\pm$0.0071), and Balona (2015; P= 4.355 d). We find a double periodicity (3.65 and 4.35 d) with both periods consistent with the previous determinations. However, at this point, it is not possible to say whether  (and which of) the periods match stellar rotation or reflect somehow the binary orbital motion (perhaps tidal locking is in play and the two periods are nearly equal?). 
KSw 85 has Hipparcos parallax of 13.52 mas corresponding to a distance of only $\sim$74 pc. We conclude that KSw 85 is a short period binary, spun up by tidal interaction leading to stellar activity, and consisting of two main sequence stars.  that is it is an RS CVn system.


\subsection {KSw 89} 

Pinsonneault et al. (2014) published the APOKASC catalog of spectroscopic and asteroseismic properties of $\sim$2000 giants in the $Kepler$ field of view. They determined $logg$=2.519$\pm$0.021 for KSw 89 from asteroseismic properties of its $Kepler$ light curve. The star was already classified as a giant by Famaey et al. (2005) on the basis of its Hipparcos distance. At a distance of about 254 pc from us KSw 89 absolute V mag is +0.36. Our spectral type and luminosity class determinations agree with these previous values.  

\subsection {KSw 91}
Reinhold et al. (2013) and Balona (2015) find matching periods (0.7186 d and 0.719 d, respectively), in close agreement to our value (0.72 d). The general agreement is that it is unlikely that this detected period is due to rotation but more likely that of a pulsating star. Pigulski et al. (2009) who conducted a parallel ground-survey to $Kepler$, classify KSw 91 as a variable with a strictly periodic sinusoidal light curve having a period of 0.71928 d. 

\section{Discussion}

Eighteen of the sample of stars discussed in this paper appear to be rapidly rotating, single subgiant and giant stars of spectral type G and K. Due to their rapid rotation, typical signatures of stellar activity are observed in their spectra (Ca II H\&K and H$\alpha$ emission) and the stars have greater than normal X-ray flux providing high f$_{x}$/f${_v}$ ratios. One star in our sample (KSw 85) is a member of a close main sequence binary, spun-up due to tidal interaction and perhaps heading to or already in the RS CVn stage and one early A-type star (KSw 69) may not exceed its usual rotation rate but is a bright X-ray source nonetheless. KSw 69 may be a rapid rotator similar to the main sequence stars Altair, Alderman, or Caph (van Belle et al., 2001, Zhao et al., 2010).  
We do not discuss these two outlier stars further and concentrate on the remaining 18 stars which we believe are currently or evolved FK Com variables. 

Typical (sub)giants have $v sin i$ values of 70, $<$25, and less than 10 km/sec for F, G, and K stars respectively (Gray 1989). Rapidly rotating ($v sin i$ values of 70-200 km/sec), apparently single, F-K (sub)giant stars are rare with only three or four confirmed FK Com variables (FK Com, ET Dra, V1794 Cyg (HD 199178), and YY Men: Jetsu et al. 1993).  This class of variable star was first hinted at by Merrill (1948) after FK Com showed a peculiar spectrum (H $\alpha$ emission) with an unusually large $v sin i$ for its spectral class. Bopp and Rucinski (1981) suggested that FK Com represents a class of single, rapidly rotating subgiant and giant stars having $v sin i$ values of 90-200 km/sec and matching a short-lived evolutionary phase described by Webbink (1976). 
Bopp and Stencel (1981) discuss this small class of G and K subgiants or giants with extreme $v sin i$ values ($\sim$100 km/sec)  providing rotation periods of a few days. They suggested that the stars represent a new class of variable, showing signs of active chromospheres and X-ray emission, but not stars that have been spun by the presence of close companions, pointing out that evolution from the main sequence to the subgiant/giant stage, {\it{without any loss of angular momentum}} would result in rotation speeds exceeding breakup. 
The FK Com variable phase includes not only an evolved single fast spinning star but the presence of an excretion disk, particularly noted in the double-peaked, variable H$\alpha$ profile. Excretion disks are uncommon in astrophysics, probably best known to exist in the  important
class of ``classical" Be stars.
The complex H$\alpha$ emission line profile has  been studied in detail by 
Ramsey et al. (1981), Huenemoerder et al. (1993), Vida et al., (2015), and Ayres et al (2016), the latter authors suggesting that the H$\alpha$ emission comes from a complex stellar magnetosphere.
Bopp and Stencel (1981) concluded that the best explanation for the FK Comae stars was to be found in the stellar evolution models of Webbink (1976) for close binary stars. 

The evolution into and out of the FK Com phase was modeled by Webbink (1976). Webbink produced a set of evolution models to examine the fate of W UMa contact binaries (EW variables) as they begin to evolve, noting that "a contact binary cannot survive as a binary beyond the main sequence", a merger must occur.  Webbink notes that descendants of W UMa binaries will appear to be single stars albeit with very rapid rotations, spun up via the merger of the two components. 
His models predict many of the observable properties of the stars, in particular the short-lived ($\sim$100 million years) FK Com phase (i.e., rare fast rotating single stars hosting an excretion disk). Webbink's models also establish the long-lived post-FK Com stages, a phase containing single spun up active evolved F-K stars lasting approximately 2 billion years until the stars are red giants near the end of double-shell burning. Webbink postulates that it is the coalescence of main sequence W UMa binaries, a process taking about 1 billion years from ZAMS turn-on to hydrogen core exhaustion and binary merger, that form the short-lived FK Com phase. Following the fast FK Com phase, the expelled material dissipates, the disk disappears, and one is left with very active, rapidly rotating, post-FK Com single stars. These now diskless stars evolve in a series of ever slowing rotation (200 km/sec to 90 km/sec), but ever expanding radius (up to 50-100 R-sun) subgiant and giant stars culminating at the red giant phase. The end product of this model sequence for the maximum expanded, double-shell burning red giant has an envelope rotation value of $\la$1 km/sec and perhaps a detached, rapidly rotating core. Such stars may be related to those red giants with rapidly rotating cores studied using $Kepler$ data by Mosser et al. (2012).

Hagan \& Stencel (1985)
performed a high-resolution spectroscopic survey of giant stars to search for large  $v sin i$ values, that is, to find FK Com-like stars. They detected no evidence for rapid rotation in 27 giants they surveyed,  concluding that such stars are rare.
We discovered eighteen rapidly rotating, apparently single stars in our X-ray survey of $\sim$6 sq. degrees of the $Kepler$ field of view. We are postulating that these stars are evolved FK Com stars, initially formed through mergers of W UMa contact binaries.
If we assume that our sample of a portion of the $Kepler$ field was typical of the entire region, there should be about 300
of these rapid rotators in the entire $\sim$100 sq. degree $Kepler$ field. 

Assuming our single rapid rotators do evolve from main sequence W UMa binaries, we can estimate the number of rapid rotators  expected.
Percy (2000) notes that 1 in every 500 main sequence F and G stars is a W UMa binary. Ciardi et al. (2011) showed that $Kepler$ observed about 80,000 dwarf stars over its entire field of view, leading to an estimated 160 W UMa binaries. 
In a steady state model at any given time, the 160 W UMas would evolve to $\sim$160 rapidly rotating single stars covering the 
evolution sequence of Webbink including the very rapid FK Com stage and the much longer timescale subgiant and giant stages. Given the rapid FK Com phase (supported by the fact that very few such stars are currently known) compared to the much longer time evolution up to the red giant phase, most of the merged rapidly rotating single stars would be, at any given time, in the subgiant and giant phases. Webbink's model would therefore suggest $\sim$160 or so such stars in the entire $Kepler$ field at any given time. Since we covered only a small fraction of the entire $Kepler$ field ($\sim$6 vs. $\sim$100 sq. degrees), we should find about 
10-15 such stars in our sample. Thus, within the framework of our simple approximation, our discovery of eighteen FK Com related stars is in line with the model prediction. 

It is interesting to note that while we found apparently single rapidly rotating stars using an X-ray selection method, other groups adopting different approaches are also reporting a number of rapidly rotating giant and subgiant stars. For example, Costa et al. (2015) analyzing the $Kepler$ light curves of 1916 giant stars classified as such on the basis of asteroseismic studies (Pinsonneault et al 2014; Tayar et al. 2015) conclude that 1.2\% of their sample matches rapidly rotating ($ v sin i$ = 10-30 km/sec) G and K giant stars. Similarly, Rodrigues da silva et al. (2015) using high resolution spectroscopy of G and K stars of class IV to Ib from the bright star catalog report that 0.8\% of those have $v sin i$ in excess of 10 km/s up to as large as 65 km/sec.
Whether these studies imply that the samples represent different type of objects (i.e. different formation mechanisms) or suggests that our unresolved stars are also rapidly rotating, cannot be said at this stage. Certainly stellar merging is a fact and now that the phenomenon has been observed and identified (Tylenda et al. 2011; Mason et al. 2010; Kamisky et al. 2015), 
Kochanek et al. (2014), on the basis of statistical and population considerations, has estimated they mergers may
be as common as 1-2 events per year. 

An interesting note here is that in the first detailed paper discussing rapidly rotating stars and 
procedures to use in order to measure their line widths (Shajn and Struve 1929) named W UMa itself as one star highly suspected to show rotationally widened spectral lines. Schilt (1927) considered W UMa as a system that had just undergone {\it fission} - the two stars breaking out of a large, single rapidly rotating main sequence star.  Only 100 years later we firmly believe W UMa binaries are soon to undergo {\it fusion} - the two stars merging into a more massive, rapidly rotating single star. 

\section{Summary}
We have presented photometric and spectroscopic observations for twenty X-ray bright stars located in the $Kepler$ field of view. 
Eighteen of the stars are evolved subgiants or giants, chosen for their large X-ray to optical flux, greater then 100 times the Sun at solar maximum and of spectral type G-K. These eighteen apparently single stars show evidence for extremely rapid rotation, X-ray bright, chromospheric activity, light curve flares. One system is a main sequence close binary and a likely RS CVn variable and one object is X-ray bright for unknown reasons.
We associate the 18 evolved stars with the objects in the evolutionary sequence put forth by Webbink (1976), starting with W UMa mergers, a rapid FK Com phase, and finally a longer-lived period as rapidly rotating single (sub)giants on their way to the red giant branch.

\acknowledgements{
We wish to thank the staff of the $Kepler$ project at NASA Ames research Center, the NASA Exoplanet Archive, and the Mt. Palomar Observatory for their continued support of the $Kepler$ mission and its follow-up work. KLS acknowledges support from the NASA Earth and Space Sciences Fellowship (NESSF).
}

{\it Facilities:} Hale 200" telescope, $Kepler$, NASA Exoplanet Archive

\begin{deluxetable}{ccccc}
\small
\tablenum{1}
\tablecolumns{5}
\tablewidth{0in}
\tablecaption{$Kepler$ Space Telescope Observing Log\tablenotemark{a}}
\tablehead{
\colhead{KSw} & \colhead{KIC} & \colhead{LC Quarters} & 
\colhead{Phot. Prec. (ppm)\tablenotemark{b}} & \colhead{SC Quarters} 
 }
 \startdata
1 & 7730305 & 0-17 & 20 & 1 \\
13 & 7732964 & 0-17 & 47 & - \\
14 & 7339348 & 1-17 & 61 & 2 \\
16 & 7505473 & 14-17 & 112 & - \\
19 & 7739728 & 0-17 & 107 & - \\
22 & 6190679 & 0-17 & 18 & - \\
28 & 7350496 & 0-17 & 2400 & 1 \\
38 & 7107762 & 0-17 & 147 & - \\
47 & 6365080 & 0-17 & 48 & - \\
54 & 7447756 & - & -  & - \\
57 & 7286410 & 0-17 & 152 & - \\
66 & 6870455 & - & - & - \\
69 & 6371741 & 2-3 & 296 & - \\ 
71 & 6372268 & 0-17 & 58 & - \\ 
73 & 6380580 & 1,2,5,6,9,10,13,14,17 & 56.5 & - \\
76 & 6224104 & - & - & - \\
78 & 4857678 & 0-10 & 7 & - \\
85 & 5557932 & 0-17 & 11 & 2 \\
89 & 6150124 & 1-17 & 7.5 & - \\
91 & 5733906 & 0-17 & 72 & 3 \\
\enddata

\tablenotetext{a}{Long cadence = 30 minute integrations, Short cadence = 1 minute integrations.}
\tablenotetext{b}{Note that 1000 ppm = 1 millimagnitude.}

\end{deluxetable}

\begin{deluxetable}{cccc}
\small
\tablenum{2}
\tablecolumns{4}
\tablewidth{0in}
\tablecaption{Hale Telescope 200" Observing Log}
\tablehead{
\colhead{KSw} & \colhead{V Mag} & \colhead{UT Date} & \colhead{Blue/Red  Int. time (secs)} \\
 }
\startdata
1 & 9.3 & 2014 Aug 27 & 200  \\ 
13 & 11.0 & 2014 Aug 27 & 300 \\
14 &  11.5 & 2014 Aug 28 & 300 \\
16 &  12.6 & 2014 Aug 27 & 700 \\
19 & 12.5 & 2014 Aug 27 & 400 \\
22 &  9.0 & 2014 Aug 28 & 200 \\
28 &  16.4 & 2014 Aug 27 & 30  \\
38 &  13.0 & 2014 Aug 27 & 500 \\
47 & 11.0 & 2014 Aug 27 & 400 \\ 
54 & 7.3 & 2014 Aug 28 & 20 \\
57 & 13.1 & 2014 Aug 28 & 500 \\
66 & 7.7 & 2014 Aug 28 & 20 \\
69 & 14.0 & 2014 Aug 28 & 600 \\
71 & 11.4 & 2014 Aug 27 & 400 \\
73 & 11.4 & 2014 Aug 28 & 400 \\
76 & 15.0 & 2014 Aug 28 & 700 \\
78 & 7.0 & 2014 Aug 28 & 20 \\
85 & 8.0 & 2014 Aug 28 & 30  \\
89 & 7.3 & 2014 Aug 28 & 20 \\
91 & 11.8 & 2014 Aug 28 & 400 \\
\enddata


\end{deluxetable}

\begin{deluxetable}{cccccccc}
\rotate
\tablenum{3}
\tablecolumns{8}
\tablewidth{8in}
\tablecaption{Description of Spectrum, Velocities, and Flaring}
\tablehead{
\colhead{KSw} & \colhead {Sp. Type/} & \colhead{Ca II H \& K em} & 
\colhead{H$\alpha$ line} & \colhead {Blue $v sin i$} & \colhead {Red $v sin i$} &
\colhead {Log(f$_{x}$/f$_{v}$)} & \colhead{Flares?} \\
\colhead{} & \colhead{Lum Class} & \colhead{} & \colhead{} & 
\colhead{km/sec $\pm$ 20\%} & \colhead{km/sec $\pm$20\%} & \colhead {} & \colhead{} \\
}
\startdata
FK Com & G4 III & yes & strong em. & 100-120 & 100-120 & -4.0 to -3.5 & yes \\
1 &  G0-2 V-IV & no  & ab. & $<$70 & $<$60 & -3.66 & no \\
13 &  G6-7 IV-III & yes & broad, complex em. & 98.4 & 107 & -2.36 & yes \\
14 &  G0-2 IV-III & no & ab. & 72.7 & $<$60 & -2.49 & yes \\
16 &  G0-2 IV-III & no & strong ab. & $<$70 & $<$60 &-2.30 & no \\
19 &  G6-7 III & yes, strong & broad, complex em. & 102.9 & 107 & -2.25  & yes \\
22 &  K2-4 IV-III & yes, strong & narrow ab. & sat. &  71.3 & -2.90 & no \\
28 &  G0-2 III & yes, weak & strong ab. & 77.5 & 64.2 & -3.60 & yes \\
47 & F7-9 V-IV & no & ab. & 82.4 & 89.6 & -2.8 & no\\
38 &  G6-7 IV-III & yes, strong & broad, complex em. & 97.3 & 105 & -1.71 & yes \\
54 &  K5-7 IV-III & yes & ab. & sat. &  82 & -4.52 & - \\
57 &  K0-3 IV-III & yes, strong & broad, strong em. & 113.0 & 113 & -1.93 & yes \\
66 &  K5-7 III & yes & ab. & sat. &  83 & -4.67  & - \\
69 & A6-9 IV-III & no & ab. & 72 & 78 & - & no\\
71 &  K0-2 IV-III  & yes, strong & ab. & 98.6 &  116 &-2.32 & yes? \\
73 &  G0-2 IV-III & yes, weak & broad ab. & $<$70 & 74.5 &-2.79 & yes?\\
76 & F8-G2 III & no & ab. & 90 & 104 & -1.2 & - \\ 
78 & F2-5 III & no & ab. & 108 & 137 & - & no\\
85 &  G0-2 V-IV  & no & ab. & $<$70 & 65.8 & -4.19 & no \\
89 &  G9-0 III  & no & ab. & 84.2 &  72.8 & -4.59 & no \\
91 & G6-7 IV-III & yes & broad, complex em. & 96.1 & 98 & -2.47 & yes \\
\enddata


\end{deluxetable}

\begin{deluxetable}{cccc}
\small
\tablenum{4}
\tablecolumns{4}
\tablewidth{4in}
\tablecaption{Measured Light Curve Periods and Calculated Radii\tablenotemark{a}}
\tablehead{
\colhead{KSw} &  \colhead{Period (d)} & \colhead{Type\tablenotemark{b}} & \colhead{R$_{*}$/R$_{\odot}$} \\
}
\startdata
1 &  4.13 & R & -- \\
13 &  2.45 & R &  5 \\
14 &  2.06 & R & 2.9 \\
16 &  0.03: &  P: & -- \\
19 & 10.74 & R & 23  \\
22 &  37.6 & R & 53 \\
28 &  10.96 & R &  15 \\
38 &  0.54 & P & - \\
47 &  2.79 & R & 4.7 \\
54 &  -- & -- & --\\
57 &  0.96 & R: & 2.1 \\
66 &  -- & -- & --\\
69 & 15: & R & 22.4 \\
71 &  5.22 & R & 10.5 \\
73 &  0.71 / 0.51 & P: & --  \\
76 & -- & -- & -- \\
78 &  0.75 & P: & - \\
85 &  3.65 / 4.35 & B & -- \\
89 &  4-6:  & R: & 7.6 \\
91 &  0.72 & P: & -- \\
\enddata
\tablenotetext{a}{Entries with ``:" after the value are uncertain.} 
\tablenotetext{b}{R=rotation, P=pulsation, B=binary}
\end{deluxetable}
\end{document}